\documentclass[11pt]{article}
\usepackage{amsfonts,latexsym}

\def\underset#1#2{\mathrel{\mathop{#2}\limits_{#1}}}
\def\overset{\stackrel}
 
\makeatletter
\@addtoreset{equation}{section}
\makeatother
\renewcommand{\theequation}{\thesection.\arabic{equation}}

\newfont{\blackb}{msbm10 scaled\magstep1}
 
\newcounter{subequation}[equation]
\makeatletter

\expandafter\let\expandafter
\reset@font\csname reset@font\endcsname

\def\subeqnarray{\arraycolsep1pt
    \def\@eqnnum\stepcounter##1{\stepcounter{subequation}%
        {\reset@font\rm(\theequation\alph{subequation})}}
\jot5mm     \eqnarray}

\def\subarray{\arraycolsep1pt
    \def\@eqnnum\stepcounter##1{\stepcounter{subequation}%
        {\reset@font\rm(\alph{subequation})}}
\jot5mm     \eqnarray}

\makeatother

\newfont{\calig}{cmsy10 scaled\magstep1}

\def\text#1{\hbox{#1}}
 
\newtheorem{theorem}{Theorem}[section]

\newtheorem{remark}{Remark}[section]
\newtheorem{corollary}{Corollary}[section]
\newtheorem{lemma}{Lemma}[section]
\newtheorem{definition}{Definition}[section]
 
\newcommand{\qed}{\hfill$\Box$}

\def\nn{\nonumber}
\def\non{\nonumber\\}
\def\be{\begin{equation}}
\def\ee{\end{equation}}
\def\ben{\begin{displaymath}}
\def\een{\end{displaymath}}
\def\baa{\begin{eqnarray}}                                                    
\def\eaa{\end{eqnarray}} 
\def\baan{\begin{eqnarray*}}
\def\eaan{\end{eqnarray*}}  
\def\ba{\begin{array}}
\def\ea{\end{array}} 
\def\a{\alpha}
\def\b{\beta}
\def\g{\gamma}

\def\ka{\kappa}
\def\l{\lambda}
\def\L{\Lambda}

\def\ph{\phi}
\def\r{\rho}
\def\s{\sigma}

\def\Th{\Theta}
\def\o{\omega}
\def\Om{\Omega}

\def\half{\textstyle{\frac12}}
\def\ie{{\it i.e.}}
\def\eg{{\it e.g.}}
\def\la{\label}
\def\Ref{\ref}
\def\c{\cite}
\def\f{\frac}
\def\p{\partial}

\def\Ref#1{(\ref{#1})}

\def\0{S}
\def\1{T}

\def\tPsi{{\widetilde{\Psi}}}
\def\tpsi{{\widetilde{\psi}}}

\def\str{{\rm str}}
\def\endo{{\rm End}}

\def\hb{{\widehat{\b}}}
\def\vm{{\vec{\mu}}}
\def\vz{{\vec{z}}}
\def\vph{\varphi}
\def\Ka{{\mathcal{K}}}
\def\wtt {{\widetilde{t}}}

\begin{document}

\begin{center}

{\huge \bf Creation operators and Bethe vectors of the \(osp(1|2)\) 
Gaudin model}\\[5mm]

{\sc P. ~P.~Kulish
\footnote{E-mail address: kulish@pdmi.ras.ru \,;
On leave of absence from Steklov Mathematical Institute, Fontanka 27,
191011, St.Petersburg, Russia.} 
and N. ~Manojlovi\'c 
\footnote{E-mail address: nmanoj@ualg.pt}\\ 
\medskip
{\it \'Area Departmental de Matem\'atica, F. C. T., Universidade do Algarve\\ 
Campus de Gambelas, 8000 Faro, Portugal}}\\ 

\vskip0.5cm 

\end{center}

\begin{abstract}
Gaudin model based on the orthosymplectic Lie superalgebra $osp(1|2)$ is\break\hfil 
studied. The eigenvectors of the $osp(1|2)$ invariant Gaudin hamiltonians are\break\hfil 
constructed by algebraic Bethe Ansatz. Corresponding creation operators 
are defined by a
recurrence relation. Furthermore, explicit solution to this
recurrence relation is found. The action of the creation operators on the 
lowest spin vector yields Bethe vectors of the model. The relation between 
the Bethe vectors and solutions to the Knizhnik-Zamolodchikov equation 
of the corresponding super-conformal field theory is established. 
\end{abstract} 


\section{Introduction} 

Classifying integrable systems solvable in the framework of the quantum 
inverse scattering method \c{F95,KS} by underlying dynamical symmetry 
algebras, one could say that the Gaudin models are the simplest ones being 
related to loop algebras and classical $r$-matrices. More sophisticated 
solvable models correspond to Yangians, quantum affine algebras, 
elliptic quantum groups, etc.  

Gaudin models \c{G,GB} are related to classical $r$-matrices, 
and the density of Gaudin hamiltonians 
\be \la{Gh}  
H^{(a)} = \sum_{b \neq a} r_{ab}(z_a - z_b) 
\ee 
coincides with the $r$-matrix. 
Condition of their commutativity $[H^{(a)} , H^{(b)} ] = 0$
is \break 
nothing else but the classical Yang-Baxter equation (YBE). 

The Gaudin models (GM) associated to classical  $r$-matrices of 
simple Lie algebras were studied in many papers 
(see \c{GB,S87,S99,J,BF,FFR,ReshVar,ST} and references therein).
The spectrum and eigenfunctions were found using different methods
(coordinate and algebraic Bethe Ansatz \c{GB,S87}, separated variables
\c{S99}, etc.). A relation to the Knizhnik-Zamolodchikov equation 
of conformal filed theory was established \c{BF,FFR,ReshVar}.
 
There are additional peculiarities of Gaudin models related to
classical $r$-matrices based on Lie superalgebras due to $Z_2$-grading 
of representation spaces and operators. The study of the 
$osp(1|2)$ invariant Gaudin model corresponding to the simplest 
non-trivial super-case of the $osp(1|2)$ invariant $r$-matrix 
\c{K85} started in \c{BM}. The spectrum of the $osp(1|2)$ invariant 
Gaudin hamiltonians $H^{(a)}$  was given, antisymmetry property of 
their eigenstates was claimed, and a two site model was connected 
with some physically interesting one (a Dicke model).  

The creation operators used in the $sl(2)$ Gaudin model coincide 
with one of the $L$-matrix entry \c{GB,S87}. However, in the 
\(osp(1|2)\) case, as we will show, the creation operators 
are complicated polynomials of the two generators $X^+(\l)$ and 
$v^+(\mu)$ of the loop superalgebra. We introduce $B$-operators 
belonging to the Borel subalgebra of the loop superalgebra 
${\cal L}(osp(1|2))$ by a recurrence relation. 
Acting on the lowest spin vector (bare vacuum) the 
$B$-operators generate exact eigenstates of the Gaudin hamiltonians 
$H^{(a)}$, provided Bethe equations are imposed on parameters of the states.
For this reason the $B$-operators are sometimes refereed to as the
creation operators and the eigenstates as the Bethe vectors, or
simply $B$-vectors. Furthermore, the recurrence relation is solved 
explicitly and the commutation relations between the 
$B$-operators and the generators of the loop superalgebra ${\cal L}(osp(1|2))$ 
as well as the generators of the global superalgebra $osp(1|2) \subset 
{\cal L}(osp(1|2))$ are calculated. We prove that the constructed 
states are lowest spin vectors of the global finite dimensional 
superalgebra $osp(1|2)$, as it is the case for many invariant 
quantum integrable models \c{TF}. Moreover, a striking coincidence
between the spectrum of the $osp(1|2)$ invariant Gaudin hamiltonians 
of spin $s$ and the spectrum of the hamiltonians of the 
$sl(2)$ Gaudin model of the integer spin $2s$ is found.

A connection between the $B$-states, when the Bethe equations are not imposed 
on their parameters, of the Gaudin models for simple Lie algebras to the solutions 
to the Knizhnik-Zamolodchikov equation was established in the papers \c{BF, FFR}. 
An explanation of this connection based on Wakimoto modules at critical level 
of the underlying affine algebra was given in \c{FFR}. An explicit form of the 
Bethe vectors in the coordinate representation was given in both papers \c{BF, FFR}. 
The coordinate Bethe Ansatz for the $B$-states of the $osp(1|2)$ Gaudin model 
is obtained in our paper as well. Using commutation relations between the 
$B$-operators and the transfer matrix $t(\l)$, as well as the hamiltonians 
$H ^{(a)}$, we give an algebraic proof of the fact that explicitly constructed 
$B$-states yield a solution to the Knizhnik-Zamolodchikov equation corresponding 
to a super-conformal field theory. This connection permits us to calculate 
the norm of the eigenstates of the Gaudin hamiltonians.
An analogous connection is expected between quantum $osp(1|2)$ spin 
system related to the graded Yang-Baxter equation \c{K85,M,STsu} and quantum 
Knizhnik-Zamolodchikov equation following the lines of \c{TV}. 
We point out possible modifications of the Gaudin hamiltonians 
and corresponding modifications of the Knizhnik-Zamolodchikov equation, 
similar to the case of the $sl(2)$ Gaudin model which was interpreted in 
\c{R92,BK} as a quantization of the Schlesinger system. 

The norm and correlation functions of the $sl(2)$ invariant 
Gaudin model were evaluated in \c{S99} using Gauss factorization 
of a group element and Riemann-Hilbert problem. The study of this 
problem for the Gaudin model based on the $osp(1|2)$ Lie superalgebra
is in progress. However, we propose a formula for the scalar products of 
the Bethe states which is analogous to the $sl(2)$ case.

The paper is organized as follows. In Section 2 we review main data 
of the quantum $osp(1|2)$ spin system: the $osp(1|2)$ invariant 
solution to the graded Yang-Baxter equation ($R$-matrix), monodromy 
matrix $T(\l)$, the transfer matrix $t(\l) = \half str T(\l)$, its 
eigenvalues and the Bethe equations. The eigenvectors of this
quantum integrable spin system can be constructed only by a complicated recurrence 
procedure \c{T} which is not given here. Nevertheless it is useful to 
remind the main data of the quantum integrable spin system because
some characteristics of the corresponding Gaudin model can be obtained
easily as a quasi-classical limit of these data. The $osp(1|2)$ Gaudin 
model and its creation operators $B_M$ are discussed thoroughly in 
Section 3. Some of the most important properties of these operators are 
formulated and demonstrated: antisymmetry with respect to their arguments, 
commutation relations with the loop
superalgebra generators,
commutation relations with the generating function $t(\l)$ of the Gaudin 
hamiltonians, a differential identity, valid in the case of the Gaudin 
realization of the loop superalgebra. Using these properties of the 
$B$-operators we prove in Section 4 that acting on 
the lowest spin vector $\Om _-$ these operators 
generate eigenvectors of the generating function 
of integrals of motion, provided the Bethe equations are 
imposed on the arguments of the $B$-operators. Possible 
modifications of the Gaudin hamiltonians are pointed out, also. 
An algebraic proof is given in Section 5 that 
constructed Bethe vectors are entering into 
solutions of the Knizhnik-Zamolodchikov equation of super-conformal 
field theory. Quasi-classical asymptotic with respect to a
parameter of the Knizhnik-Zamolodchikov equation permits us
to calculate the norm of the eigenstates of the Gaudin hamiltonian.
Further development on possible evaluation of correlation functions 
is discussed in Conclusion. 

  
\section{$OSp(1|2)$-invariant $R$-matrix}   

Many properties of the Gaudin models can be obtained as a quasi-classical 
limit of the corresponding quantum spin systems related to solutions to 
the Yang-Baxter equation
\be \la{YBE}  
R_{12}(\l - \mu) R_{13}(\l - \nu) R_{23}(\mu - \nu) 
=  R_{23}(\mu - \nu) R_{13}(\l - \nu) R_{12}(\l - \mu)  \,,
\ee
here the standard notation of the quantum inverse scattering method \c{F95,KS} 
is used to denote spaces $V_j, j=1, 2, 3$ on which 
corresponding $R$-matrices $R_{ij}, ij = 12, 13, 23$ act non-trivially. 
In the quasi-classical limit $\eta \to 0$
\ben 
R(\l; \eta) = I + \eta r(\l) + {\cal O}(\eta^2), 
\een
some relations simplify and therefore can be solved explicitly providing 
more detailed results for the Gaudin model. 

The graded Yang-Baxter equation \c{KS, K85} differs from 
the usual Yang-Baxter equation \Ref{YBE} by some sign factors 
due to the embedding of $R$-matrix into the space of matrices 
acting on the $Z_2$-graded tensor product $V_1 \otimes V_2 \otimes V_3$. 
At this point our aim is to reach fundamental $osp(1|2)$ invariant solution.
The rank of the orthosymplectic Lie algebra $osp(1|2)$
is one and its dimension is five. The three even generators are $h, X^+, X^-$ and 
the two odd generators are $v^+, v^-$ \c{Ritt}. The (graded) commutation relations 
between the generators are 
\be \la{osp}
\ba{cc}
\left[ h, X^{\pm} \right] = \pm 2 X^{\pm} \;,&
\left[ X^+, X^- \right] = \, h  \;, \\ 
\left[ h, v^{\pm} \right] = \pm v^{\pm} \;, &  
\left[ v^+, v^- \right] _+ = - h \;, \\
\left[ X^{\mp}, v^{\pm} \right] = \, v^{\mp} \;, &  
\left[ v^{\pm}, v^{\pm} \right] _+ = {\pm} 2 X^{\pm} \;, 
\ea
\ee
together with $\left[ X^{\pm}, v^{\pm} \right] = 0$. Notice that the 
generators $h$ and $v^{\pm}$ considered here \Ref{osp} differ by 
a factor of 2 from the ones used in \c{Ritt,K85}. Thus the Casimir 
element is 
\baa \la{Cas}  
c_2 &=& h^2 + 2 \left(X^+ X^- + X^- X^+ \right) 
+ \left(v^+ v^- - v^- v^+ \right)  \nonumber\\
&=& h^2 - h + 4 X^+ X^- + 2 v^+ v^-  \,.
\eaa
For further comparison with the $sl(2)$ Gaudin model \c{GB,S87} 
and due to the chosen set of generators \Ref{osp} we parameterize 
the finite dimensional irreducible representations $V^{(l)}$ of the 
$osp(1|2)$ Lie superalgebra by an integer $l$, so that their dimensions 
$2l + 1$ and the values of the Casimir element \Ref{Cas} $c_2 = l(l+1)$ 
coincide with the same characteristics of the integer spin $l$ 
irreducible representations of $sl(2)$. 

We proceed to write down the $osp(1|2)$ invariant solution of the graded 
Yang-Baxter equation and the main data of the corresponding quantum 
spin system: the $L$-operator, the transfer matrix $t(\l)$, 
the eigenvalue of the generating function of the integrals of 
motion $t(\l)$ and the Bethe equations. 

The fundamental irreducible representation $V$ of $osp(1|2)$ 
is three dimensional. We choose a gradation (parity) 
of the basis vectors $e_j;\, j=1,2,3$ to be $(0, 1, 0)$. 

The invariant $R$-matrix is a linear combination \c{K85} 
\be \la{R} 
R= \l \left( \l+\frac {3\eta}2 \right )I + \eta
\left( \l+\frac {3\eta}2 \right) {\cal P} - \eta \, \l \, K \,, 
\ee 
of the three $OSp(1|2)$ group invariant operators 
\be \la{inv}
[ g\otimes g, X] = 0\,, \quad g \in OSp(1|2), \quad  
X \in {\endo} \, (V \otimes V),  
\ee
acting on $V \otimes V$: the identity $I$, the permutation $\cal P$ 
and a rank one projector $K$. In the equation \Ref{R} $\l$ is the spectral 
parameter, and $\eta$ is a quasi-classical parameter. 

Using the projectors on the irreducible representation components in 
the Clebsch-Gordan decomposition
$V\otimes V = V ^{(2)}\oplus V ^{(1)}\oplus V ^{(0)}$
\be \la{proj}
P_2 = \f 12 \left( I + {\cal P} \right) \;, \quad
P_1 = \f 12 \left( I - {\cal P} - 2 K \right) \;, \quad
P_0 = K \;,
\ee 
one can represent the $R$-matrix \Ref{R} in the form
\be \la{R-proj}
R (\l, \eta) = \left( \l+\eta \right) \left(\l+ \f {3\eta}2 \right)
\left( P_2 + \f{\l-\eta}{\l+\eta} P_1 + \f {\l-3\eta/2}{\l+3\eta/2} 
P_0 \right)  
\ee 
Deforming these projectors to the quantum superalgebra $osp_q(1|2)$
projectors \c{KR,HS} $P_i \to P _i (q)$ and substituting rational 
functions by trigonometric ones $( \l \pm\eta ) \to \sinh ( \l \pm\eta )$,
one arrives to a trigonometric solution to the graded Yang-Baxter 
equation, in the braid group form, and corresponding 
anisotropic models \c{HS,LS}.
 
The $L$-operator of the quantum spin system on a one-dimensional lattice 
with $N$ sites coincides with $R$-matrix acting on a tensor product 
$V_0 \otimes V_a$ of auxiliary space $V_0$  and 
the space of states at site $a = 1, 2,\dots N$ 
\be \la{Lq} 
L_{0a}(\l - z_a) = R_{0a} (\l - z_a) \,, 
\ee 
where \(z_a\) is a parameter of inhomogeneity (site dependence) \c{F95,KS}. 
Corresponding monodromy matrix $T$ is an ordered product of the $L$-operators 
\be \la{Tq} 
T(\l ; \{z_a\}_1^N) = L_{0N}(\l - z_N) \dots  
L_{01}(\l - z_1) = \underset{\longleftarrow}{\prod_{a=1}^{N}}
L_{0a} (\l - z_a) \,.  
\ee
The commutation relations of the $T$-matrix entries follow from the 
FRT-relation \c{F95} 
\be \la{RTT}  
R_{12}(\l - \mu)T_1(\l)T_2(\mu) = 
T_2(\mu)T_1(\l)R_{12}(\l - \mu) \,.  
\ee 
Multiplying \Ref{RTT} by $R_{12}^{-1}$ and taking the super-trace 
over $V_1 \otimes V_2$, one gets commutativity of the transfer 
matrix 
\be \la{tq} 
t(\l) = \sum_j (-1)^{j+1}T_{jj}(\l ; \{z_a\}_1^N)= T_{11} - T_{22} + T_{33}  
\ee 
for different values of the spectral parameter 
$t(\l)t(\mu) = t(\mu)t(\l)$.

The choice of the $L$-operators \Ref{Lq} corresponds to 
the following space of states of the $osp(1|2)$-spin system  
\ben
{\cal H} = \underset {a=1}{\overset{N}{\otimes}} V_a \;. 
\een
The eigenvalue of the transfer matrix $t(\l)$ 
in this space is \c{K85} 
\baa 
\la{lq} 
\L (\l ; \{\mu_j\}_1^M) &=& 
\a_1^{(N)}(\l;\{z_a\}_1^N) \prod_{j=1}^M S_1(\l - \mu_j) 
-\a_2^{(N)}(\l; \{z_a\}_1^N) \times
\non
&\times& \prod_{j=1}^M S_1 \left( \l - \mu_j +\f {\eta}2 \right) 
S_{-1}(\l - \mu_j +\eta) + 
\non
&+&  \a _3^{(N)}(\l; \{z_a\}_1^N) \prod_{j=1}^M 
S_{-1} \left( \l - \mu_j + \f {3\eta}2 \right) \,, 
\eaa
where $\a_j^{(N)}(\l;\{z_a\}_1^N)=\prod_{b=1}^N 
\a_j(\l - z_b)\,; j = 1, 2, 3\,,$  
\baa
\a_1(\l) &=& \left( \l+ \eta \right) \left( \l+ 3\eta/2\right) \,,   
\quad \a_2(\l) = \l  \left( \l + 3\eta/2 \right) \,,  
\non
\a_3(\l) &=& \l \left( \l + \eta/2 \right) \,,  
\quad S_n(\mu) = \f {\mu - n\eta/2 }{\mu + n\eta/2 } \;. 
\eaa 
Although according to \Ref{lq} the eigenvalue has formally two 
sets of poles at $\l = \mu_j - \eta/2$ and $\l = \mu_j - \eta$, the 
corresponding residues are zero due to the Bethe equations \c{K85} 
\be \la{BEq} 
\prod_{a=1}^N
\left(\f{\mu_j -z_a +  \eta/2}{\mu_j -z_a - \eta/2}\right) = 
\prod_{k=1}^M S_1(\mu_j - \mu_k) S_{-2}(\mu_j - \mu_k) \;.
\ee

If we take different spins $l_a$ at different sites of the lattice 
and the following space of states 
\ben
{\cal H} = \underset {a=1}{\overset{N}{\otimes}} V^{(l_a)}_a \;, 
\een
then the factors on the left hand side of \Ref{BEq} will 
be spin dependent too
\ben
\f{\mu_j -z_a +  \eta l_a/2}{\mu_j -z_a - \eta l_a/2}\;. 
\een

The \(osp(1|2)\) invariant $R$-matrix \Ref{R} has more complicated 
structure than the $sl(2)$ invariant $R$-matrix of C. N. Yang
$R= \l I + \eta {\cal P}$. As a consequence the commutation relations of 
the entries $T_{ij}(\l)$  of the $T$-matrix \Ref{Tq} are more complicated 
and construction of the eigenstates of the transfer matrix $t(\l)$ by 
the algebraic Bethe Ansatz can be done only using a complicated 
recurrence relation expressed in terms of $T_{ij}(\mu_k)$ \c{T} 
(see also \c{M} for the case of $osp(1|2)$). It will be shown 
below that due to a simplification of this relation in the quasi-classical 
limit $\eta \to 0$ one can solve it and find the creation operators 
for the $osp(1|2)$ Gaudin model explicitly. Furthermore, the commutation 
relations between the creation operators and the generators of the 
loop superalgebra as well as the generating function $t(\l)$ 
of the Gaudin hamiltonians will be given explicitly.


\section{$OSp(1|2)$ Gaudin model and corresponding creation operators}  

As in the case of any simple Lie algebra, the classical $r$-matrix of the 
orthosymplectic Lie algebra $osp(1|2)$ can be expressed in a pure algebraic 
form using (reduced) Casimir element in the tensor product 
$osp(1|2) \otimes osp(1|2)$ 
\c{K85} 
\baa \la{cra}
&\quad& {\hat r} \left( \l \right) = \frac 1{\l} 
c_2^{\otimes}  \,, \non 
&\!& c_2^{\otimes} = h \otimes h + 2 
\left( X^+ \otimes X^- + X^- \otimes X^+ \right) 
+ \left(v^+ \otimes v^- - v^- \otimes v^+ \right) \,.
\eaa

The matrix form of the Casimir element $\hat r$ in the fundamental representation 
$\pi$ of $osp(1|2)$ follows from \Ref{cra} by substituting appropriate $3 \times 3$ 
matrices instead of the $osp(1|2)$ generators (\ref{osp}) \c{Ritt} and taking 
into account $Z_2$-graded tensor product of even and odd matrices \c{K85}.
Alternatively, the same matrix form of $\hat r$ can be obtained 
as a term linear in $\eta$ in the quasi-classical expansion of 
\Ref{R}, \Ref{R-proj} 
\ben 
r (\l) = \frac 1{\l}  \left( {\cal P} - K \right) \,,  
\een
where $\cal P$ is a graded permutation matrix and $K$ is a rank one projector. 
Let us write explicitly the matrix form of $r _0 =  {\cal P} - K$ 
in the basis $e_1\otimes e_1, e_1\otimes e_2, e_1\otimes e_3, \ldots , 
e_3\otimes e_3$ of the tensor product of two copies of the fundamental 
representation $V\otimes V$
\be \la{r0}  
r_0 = \l \, r (\l) = {\cal P} - K = \left( \ba{ccccccccc}
1 &  & & & & &  & & \\
&0&  & 1 & & &  & & \\
& &-1& &-1 & & 2& & \\
&1&  &0&   & &  & & \\
& & 1& & 0 & &-1& & \\
& &  & &   &0&  &1& \\
& & 2& & 1 & &-1& & \\
& &  & &   &1&  &0& \\
& &  & &   & &  & &1 
\ea \right) 
\ee 
with all the other entries of this $9\times 9$ matrix being
identically equal to zero. 

A quasi-classical limit  $\eta \to 0$ 
of the FRT-relations \Ref{RTT} results in a matrix form of the loop 
superalgebra relation \(( T(\l; \eta) = I + \eta L(\l) + {\cal O}(\eta^2))\)
\be \la{rL}  
\left[ \underset 1 {L}(\l), \, \underset 2 {L} (\mu) \right] = - 
\left[ r_{12}(\l - \mu) \, , \, \underset 1 {L}(\l) + \underset 2 {L}(\mu) \right] \,. 
\ee
Both sides of this relation have the usual commutators of even 
$9 \times 9$ matrices $\underset 1 {L}(\l)  = L(\l) \otimes I_3$,
$\underset 2 {L}(\mu)  = I_3 \otimes L(\mu)$ and $r_{12}(\l - \mu)$, where $I_3$ is 
$3 \times 3$ unit matrix and $L(\l)$ 
has loop superalgebra valued entries: 
\be \la{L}  
L(\l) = \left( \ba{ccc}
h(\l) & - v ^-(\l) & 2 X ^{-}(\l) \\
v ^+(\l) & 0 & v ^-(\l) \\
2 X ^+(\l) & v ^+(\l) & - h(\l) 
\ea \right) 
\ee 
The relation \Ref{rL} is a compact matrix form of the following commutation 
relations between the generators $h(\l), \, v^{\pm}(\mu), \, X^{\pm}(\nu)$ 
of the loop superalgebra under consideration
\baa \la{la}
\left[h(\l) \; , \; X^{\pm}(\mu) \right] = 
\mp \, 2 \frac {X^{\pm}(\l) - X^{\pm}(\mu)}{\l - \mu} \; & & \quad
\left[h(\l) \; , \; v^{\pm}(\mu) \right] = \phantom{2}
\mp \; \frac {v^{\pm}(\l) - v^{\pm}(\mu)}{\l - \mu} \,, \non 
\left[X ^+(\l) \; , \; X ^-(\mu) \right] = - \phantom{2}
\frac {h(\l) - h(\mu)}{\l - \mu}   \; & & \quad 
\left[X^{\pm}(\l) \; , \; v^{\mp}(\mu) \right] = - \phantom{2}
\frac {v^{\pm}(\l) - v^{\pm}(\mu)}{\l - \mu} \,, \non 
\left[v^+(\l)\; , \; v^-(\mu) \right]_+ =  \phantom {-} \phantom{2}
\frac {h(\l) - h(\mu)}{\l - \mu}  \;&&\quad 
\left[v^{\pm}(\l)\; , \; v^{\pm}(\mu) \right]_+ = 
\mp \, 2 \frac{X^{\pm}(\l) - X^{\pm}(\mu)}{\l - \mu}\,,\non 
\eaa
together with \( \left[X^{\pm}(\l) \; , \; v^{\pm}(\mu) \right] = 0 \).

Actually these commutation relations \Ref{la} define the positive part 
${\cal L}_+(osp(1|2))$ of the loop superalgebra. The usual generators $Y_n$  
of a loop algebra parameterized by non-negative integer, 
are obtained from the expansion of $Y(\l)$ 
\ben
Y(\l) = \sum_{n \ge 0} \f {Y_n}{\l^{n+1}} . 
\een 
In particular, taking all $Y_n = 0$ for $n > 0$ one gets an $L$-operator 
$L(\l) = L_0/\l$, where $L_0$ is $osp(1|2)$-valued matrix. This $L_0$ 
similar to $r_0$ satisfies  cubic characteristic equation
with the $osp(1|2)$ Casimir element \Ref{Cas} as coefficient
\baa \la{L0-ev}
L ^3_0 + 2 L ^2_0 - ( c _2 - 1 ) \, L_0 - c _2 I = 0 \;.
\eaa

A Gaudin realization of the loop algebra \Ref{la} can be defined
through the generators $Y = (h, v^{\pm}, X^{\pm})$ 
\baa \la{Gr}  
Y(\l) = \sum_{a=1}^N \frac{Y_a}{\l - z_a} \;, \quad 
Y_a \in {\endo} \, (V_a)\,, 
\eaa
where $Y_a$ are \(osp (1|2) \) generators in an irreducible 
representation $V^{(l _a)}_a$ of the lowest spin $-l _a$ 
associated with each site $a$ \c{GB,S87}. Then the $L$-operator 
\Ref{L} has the form  
\baa \la{LG}  
L(\l; \{z_a\}_1^N) &=& \sum_{a=1}^N \frac{L_a}{\l - z_a}\,,  
\eaa
here $\{z_a\}_1^N$ are parameters of the model (cf \Ref{Lq}, \Ref{Tq}). 
It follows from the relation \Ref{LG} that the first term in the asymptotic
expansion near $\l = \infty$ defines generators of the global 
$osp(1|2) \subset {\cal L} _+ (osp(1|2))$ 
\baa \la{Lgl}  
L_{gl} = \lim _{\l \to \infty} \l \, L(\l) &=& \sum_{a=1}^N L_a \,,  
\eaa
where
\be \la{Lgl-m}  
L_{gl} = \left( \ba{ccc}
h_{gl} & - v ^-_{gl} & 2 X ^-_{gl} \\
v ^+_{gl} & 0 & v ^-_{gl} \\
2 X ^+_{gl} & v ^+_{gl} & - h_{gl} 
\ea \right) \;.
\ee 
Moreover, from the equation (\ref{rL}) we get
\baa \la{rLgl}  
\left[ \underset 1 {L_{gl}}, \, \underset 2 {L} (\mu) \right] = - 
\left[ r_0 \, , \, \underset 2 {L}(\mu) \right] \,, 
\eaa
here \( \underset 1 {L_{gl}} = L_{gl} \otimes I_3\), 
\eg $[h_{gl} \, , \, v ^+ (\mu) ] =  v ^+ (\mu)$.

Let us consider the loop superalgebra ${\cal L}_+(osp(1|2))$ as the
dynamical symmetry algebra, {\ie} as the algebra of observables. 
In order to define a dynamical system besides the algebra of observables
we need to specify a hamiltonian. It is a well-known fact that due 
to the $r$-matrix relation \Ref{rL}, the so-called  Sklyanin linear brackets, 
the elements 
\baa\la{tG} 
t(\l) &=& \frac 12 \; {\str} \; L^2(\l) = h^2(\l) + 
2 [X^+(\l)\,,\,X^-(\l)]_+ + [v^+(\l)\,,\,v^-(\l)]_- 
\nonumber\\ 
&=&h^2(\l) + h'(\l) + 4 X^+(\l)X^-(\l) + 2 v^+(\l)v^-(\l) 
\eaa 
commute for different values of the spectral parameter 
$t(\lambda)t(\mu) = t(\mu)t(\lambda)\,.$ Thus, $t(\lambda)$ can be 
considered as a generating function of integrals of motion. 

It is straightforward to calculate the commutation relations
between the operator $t(\l)$ and the generators of the loop algebra 
$X^+(\mu )$ and $v^+(\mu)$
\be \la{tX}   
\left [ t(\l), X^+(\mu) \right] = 4 
\f {X^+(\mu)h(\l) - X^+(\l)h(\mu)}{\l - \mu} 
- \phantom{2}\f {v^+(\l)v^+(\mu) - v^+(\mu)v^+(\l)}{\l - \mu} \;,
\ee
\be \la{tv}   
\left[ t(\l), v^+(\mu) \right] = 2 
\f {v^+(\mu)h(\l) - v^+(\l)h(\mu)}{\l - \mu} 
- 4 \f {X^+(\mu)v^-(\l) - X^+(\l)v^-(\mu)}{\l - \mu} \;.
\ee
These relations will be important for the proof of the lemma 3.6
as well as the proof of the main theorem.

A direct consequence of the equation \Ref{rLgl} is an invariance of
the generating function of integrals of motion $t(\l)$ under the 
action of the global $osp(1|2)$
\be \la{gl-sym}
\left[ t(\l), L_{gl} \right] = 0 \,.
\ee

Preserving some generality we can consider the representation space
\({\cal H}_{ph}\) of the dynamical algebra to be a lowest spin 
\(\rho (\l)\) representation of the loop superalgebra with the 
lowest spin vector \(\Omega_-\) 
\be 
\la{vac} 
h (\l) \Om_- = \rho (\l) \Om_- \,, \quad v^-(\l) \Om_- = 0 \,. 
\ee 
In particular, a representation of the Gaudin realization \Ref{LG} can be 
obtained by considering irreducible representations $V_a^{(l_a)}$ of the 
Lie superalgebra \(osp(1|2)\) defined by a spin $-l_a$ and a lowest spin 
vector $\o_a$ such that $v_a^-\o_a = 0$ 
and $h_a \o_a = -l_a \o_a$. Thus,
\be \la{Gre}
\Om _- = \mathop{\otimes} \limits_{a=1}\limits^N \o _a \;, \quad 
\hbox {and} \quad 
\rho (\l) = \sum_{a=1}^N \frac{-l_a}{\l - z_a} \;.
\ee

It is a well-known fact in the theory of Gaudin models \c{GB,S87}, 
that the Gaudin\break\hfil 
hamiltonian 
\baa \la{sGh}
H^{(a)} &=& \sum_{b \neq a} \frac {c_2^{\otimes}(a,b)}{z_a - z_b} \non
&=& \sum_{b \neq a} \frac 1{z_a - z_b}
\left(h _a h _b + 2 \left( X ^+_a X ^-_b + X ^-_a X ^+_b \right) 
+ \left( v  ^+_a v ^-_b - v ^-_a v ^+_b \right)\right) \non
\eaa
can be obtained as the residue of the operator $t(\l)$ at the point $\l = z_a$ 
using the expansion 
\baa \la{tpole}
t(\l) &=& \sum_{a =1}^N \left( \frac {l_a(l_a+1)}{(\l - z_a)^2} + \, 2 \,
\frac {H^{(a)}}{\l - z_a} \right)  \, .
\eaa

To construct the set of eigenstates of the generating function 
of integrals of motion $t(\l)$ we have to define appropriate 
creation operators. The creation operators used in the $sl(2)$ 
Gaudin model coincide with one of the $L$-matrix entry \c{GB,S87}. 
However, in the \(osp(1|2)\) case the creation operators are 
complicated functions of the two generators of the loop superalgebra 
$X^+(\l)$ and $v^+(\mu)$.

\begin{definition}
Let $B_{M}(\mu_1, \dots , \mu_M)$ belong to the Borel subalgebra $\cal B$
of the \(osp (1|2) \) loop algebra ${\cal L}_+ \left(osp (1|2)\right)$
such that
\be \la{rr}  
B_{M}(\mu_1, \dots , \mu_M) = v^+(\mu_1) B_{M-1}(\mu_2, \dots , \mu_M) 
+ 2 X^+(\mu_1) \sum_{j=2}^{M} 
\f {(-1)^j}{\mu_1 - \mu_j} B_{M-2}^{(j)}(\mu_2, \dots , \mu_M)  \,, \non
\ee
with $B_0 = 1,$ $B_1(\mu) = v^+(\mu) $ and $B_M = 0$ for $M < 0$. The notation 
adopted here is that upper index of $B_{M-2}^{(j)}(\mu_2, \dots ,\mu_M)$ 
means that the argument $\mu_j$ is omitted.
\end{definition}

\begin{remark} 
It may be useful to write down the explicit expressions for the first
few creation operators
\ben
B_0 = 1 \; , B _1(\mu) = v ^+ (\mu) \; , \quad
B_2(\mu_1, \mu_2) = v^+(\mu_1) v^+(\mu_2) + 2 
\f {X^+(\mu_1)}{\mu_1 - \mu_2} \; , 
\een
\be \la{fewBs}
B_3(\mu_1, \mu_2, \mu_3) = v^+(\mu_1) B_2(\mu_2, \mu_3) + 2 X^+(\mu_1) 
\left( \f {v^+(\mu_3)}{\mu_1 - \mu_2} - \f {v^+(\mu_2)}{\mu_1 - \mu_3} \right) 
\;, 
\ee
\baan
B_4(\mu_1, \mu_2, \mu_3, \mu_4) &=& v^+(\mu_1) B_3(\mu_2, \mu_3, \mu_4) 
+ 2 X^+(\mu_1) \times \\
&& \times \left( \f {B_2(\mu_3, \mu_4)}{\mu_1 - \mu_2} 
- \f {B_2(\mu_2, \mu_4)}{\mu_1 - \mu_3} 
+ \f {B_2(\mu_2, \mu_3)}{\mu_1 - \mu_4} \right) \nn
\eaan
\end{remark}

The creation operators $B_{M}(\mu_1, \dots , \mu_M)$ together with $h(\nu)$ 
generate the Borel\break\hfil 
subalgebra $\cal B \subset {\cal L}_+$. As we will show 
below, the $B$-operators are such that the Bethe vectors are generated by their 
action on the lowest spin vector $\Om _-$ \Ref{vac}. To prove this result we will need
some important properties of the $B$-operators which are summarized in the following
seven lemmas.

\begin{lemma}
The creation operators $B_{M}(\mu_1, \dots , \mu_M)$ are antisymmetric 
functions of their arguments
\baa \la{asym} 
B_{M}(\mu_1, \dots , \mu_{k},  \mu_{k+1}, \dots , \mu_M) &=& 
- \; B_{M} (\mu_1, \dots , \mu_{k+1}, \mu_{k}, \dots , \mu_M) \,,
\eaa
here $1\leq k < M$ and $M \ge 2$. 
\end{lemma}

Sometimes an alternative formulation of the recurrence relation 
\Ref{rr} can be useful. 

\begin{lemma}
Alternatively, the recurrence relation \Ref{rr} can be written 
in the following form 
\baa \la{rrop} 
B_{M}(\mu_1, \dots , \mu_M)
&=& B_{M-1}(\mu_1, \dots, \mu_{M-1}) v^+(\mu_M)
+ 2 \sum_{j=1}^{M-1} (-1)^{M-j-1} \times \non
&&\quad \times \f { X^+(\mu_j)}{\mu_j - \mu_M} \;
B_{M-2}^{(j)}(\mu_1, \dots, \mu_{M-1})  \;,
\eaa
with $B_0 = 1,$ $B_1(\mu) = v^+(\mu) $ and $B_N = 0,$ for $N < 0$.
\end{lemma}

In the subsequent lemmas we will calculate the commutation relations
between the generators of the loop superalgebra ${\cal L}(osp(1|2))$ and 
the $B$-operators. In order to simplify the formulas we will omit the
arguments and denote the creation operator $B _M (\mu_1, \dots , \mu_M)$ 
by $B _M$. We suppose that it will not be difficult to restore the\break\hfil 
appropriate set of arguments.

\begin{lemma} 
The commutation relations between the generator $v^+(\l)$ of the loop 
superalgebra and the creation operator $B_M$ are given by 
\baa \la{vB} 
v^+(\l) B_{M} &=& (-1)^M B_{M} v^+(\l) + 2 \sum_{j=1}^M (-1)^j
\f {X^+(\l) -X^+(\mu _j)}{\l - \mu _j} \; B_{M-1}^{(j)} \,,
\eaa
here as in the previous lemma the upper index of $B_{M-1}^{(j)}$ 
means that the argument $\mu_j$ is omitted.
\end{lemma}

\begin{lemma} 
The generator $h(\nu)$ has the following commutation relation with the
$B_M$ elements  
\baa \la{hB} 
h(\l) B_{M} &=&  B_{M} \left( h(\l) + \sum_{i=1}^M \frac 1{\l - \mu_i} \right)
\nonumber\\ 
&+& \sum _{i=1}^M \frac{(-1)^i}{\l - \mu _i} \left( v ^+(\l) B_{M-1}^{(i)}
+ 2 X ^+ (\l) \mathop{\sum_{j=1}} \limits_{j \neq i}\limits^M 
\f{(-1)^{j+\Th (i-j)}}{\mu _i -\mu _j} B_{M-2}^{(i,j)} \right) \;, \non
\eaa
here $\Th (j)$ is Heaviside function
\[ 
\Th (j) = \left\{ \ba{ll}
                1 & \mbox {if $j > 0$} \\
		0 & \mbox {if $j \leq 0$}
		 \ea \right. \;,
\]
and the meaning of the upper indices of $B_{M-2}^{(i,j)}$ 
is that the parameters $\mu_i, \mu_j$ are omitted. 
\end{lemma}

\begin{lemma}
The generator $v ^-(\l)$ of the loop superalgebra has the following 
commutation relation with the $B$-operators 
\baa \la{v-B}
&&v ^-(\l)  B_{M} = (-1) ^M B_{M} v ^-(\l) + \sum _{j=1}^M (-1)^{j-1} 
B_{M-1}^{(j)} \left( \f{h (\l) - h (\mu _j)}{\l - \mu _j} \right.
\non
&&+\left.\underset{k \neq j}{\sum_{k=1}^M}
\f{1}{(\l - \mu _k)(\mu _k - \mu _j)} \right )  
+ v ^+(\l) \underset{i < j}{\sum_{i,j=1}^M}
(-1) ^{i-j-1} \f {B^{(i,j)}_{M-2}}{\mu _i - \mu _j}  
\left( \f 1{\l - \mu _i} + \f 1{\l - \mu _j} \right) \; . \non
\end{eqnarray}
\end{lemma}

Already at this point we can make some useful observations.
\begin{remark}
The commutation relations between the generators of the global \(osp (1|2) \) 
\Ref{Lgl-m} and the $B_M$ elements follow from the lemmas 3.3--5. To see this 
we multiply \Ref{vB}, \Ref{hB} and \Ref{v-B} by $\l$ and then take the limit 
$\l \to \infty$. In this way we obtain 
\baa \la{vglB} 
v^+_{gl} B_{M} &=& (-1)^M B_{M} v^+_{gl} - 2 \sum_{j=1}^M (-1)^j
X^+(\mu _j) B_{M-1}^{(j)} \;, \\
\la{hglB}
h _{gl}  B_{M} &=& B_{M} \left( h _{gl}  + M \right) \;, \\
\la{v-glB} 
v^- _{gl} B_{M} &=& (-1)^M  B_{M} \, v^-_{gl} + \sum_{j=1}^M (-1)^j B_{M-1}^{(j)} 
\left( h(\mu_j) + \sum_{k \neq j}^M \f 1{\mu_j - \mu_k} \right) \,. 
\eaa 
\end{remark}

The subsequent lemma is one of the most important results. The proof 
of the main theorem is based on the following. 
\begin{lemma}
The generating function of integrals of motion \(t (\l)\) \Ref{tG} 
has the following commutation relation with the creation operator 
$B_M (\mu_1, \dots , \mu_M)$
\baa \la{tB}
t(\l)B_{M} &=& B_{M} t(\l) + 2 B_{M} \left( h (\l) \sum _{i=1}^M 
\f 1{\l -\mu _i} + \mathop{\sum_{i,j=1}} \limits_{i < j}^M 
\f 1{(\l -\mu _i)(\l -\mu _j)} \right)
\non
&+& 2 \sum _{i=1}^M \f {(-1)^i}{\l - \mu _i} \left( v ^+(\l) B_{M-1}^{(i)}
+ 2 X ^+ (\l) \mathop{\sum_{j=1}} \limits_{j\neq i}^M 
\f {(-1)^{j+\Th (i-j)}}{\mu _i -\mu _j} 
B_{M-2}^{(i,j)} \right) \hb _M (\mu _i) 
\non
&+& 4 \sum _{i=1}^M \frac{(-1)^i}{\l - \mu _i} B_{M-1}^{(i)} \left( 
X ^+ ( \l )  v ^- ( \mu _i ) - X ^+ ( \mu _i )  v ^- ( \l ) \right) \;.
\eaa
Here we use the following notation for the operator 
\[\hb _M (\mu _i) = h(\mu_i) + \mathop{\sum_{j=1}} \limits_{j\neq i}^M 
\f 1{\mu_i - \mu_j} \, . \]
\end{lemma}

In the Gaudin realization \Ref{Gr} the creation operators $B_M (\mu_1, \dots , \mu_M)$
have some specific analytical properties.
\begin{lemma}
The $B$-operators in the Gaudin realization \Ref{Gr} satisfy an important\break\hfil 
differential identity
\baa \la{derB}
\f {\p}{\p z_a} B _M = \sum _{j=1}^M  \f {\p}{\p \mu _j} \left( 
\f {(-1) ^j}{\mu_j - z_a} \left( v ^+_a B^{(j)}_{M-1} + 2 \, X ^+_a \,
\sum _{k\neq j}^M \f {(-1) ^{k+\Th (j-k)}}{\mu_j - \mu _k}  B^{(j,k)}_{M-2} 
\right) \right) \, . 
\eaa
\end{lemma}
This identity will be fundamental step in establishing a connection
between the Bethe vectors and the Knizhnik-Zamolodchikov equation.

The proofs of the lemmas are based on the induction method. 
As illustrations, we prove explicitly lemma 3.1 and lemma 3.3. 

{\it Proof of lemma 3.1.} We prove the lemma 3.1 by induction. Consider $M=2$
\ben 
B_2(\mu_1, \mu_2) = v^+(\mu_1) v^+(\mu_2) + 2 
\f {X^+(\mu_1)}{\mu_1 - \mu_2} \;.
\een
Using the commutation relations \Ref{la} it is straightforward to
check that $B_2(\mu_1, \mu_2)$ is antisymmetric
\ben 
B_2(\mu_1, \mu_2) = - B_2(\mu_2 , \mu_1) \;.
\een
Assume $B_{N}(\mu_1, \dots , \mu_N)$ is antisymmetric
for $M > N \geq 2$. We have to prove that 
$B_{M}(\mu_1, \dots , \mu_M)$ is antisymmetric also. 

Consider $j \geq 2$, the antisymmetry of $B_{M}(\mu_1, \dots , \mu_M)$
with respect to $\mu _j$ and $\mu _{j+1}$ follows directly form the
recurrence relation \Ref{rr} and our assumption. Namely, the terms 
$B_{M-2}^{(j)}(\mu_2, ...,\mu_M)/(\mu_1- \mu_j)$ and 
$B_{M-2}^{(j+1)}(\mu_2, ...,\mu_M)/(\mu_1- \mu_{j+1})$ enter with 
the opposite sign. 

Therefore we only have to show the antisymmetry of $B_{M}(\mu_1, \dots , \mu_M)$
with respect to the interchange of $\mu _1$ and $\mu _2$. To see this
we have to iterate the recurrence relation \Ref{rr} twice and combine
the appropriate terms
\baa \la{asym12}
B_{M}(\mu_1, \dots , \mu_M) &=& \left( v^+(\mu_1) v^+(\mu_2) + 2 
\f {X^+(\mu_1)}{\mu_1 - \mu_2} \right) B_{M-2}(\mu_3, \dots , \mu_M) \non 
&+& 2  v^+(\mu_1) X^+(\mu_2) \sum_{j=3}^{M} \f {(-1)^{j+1}}{\mu_2 - \mu_j} 
B_{M-3}^{(j)}(\mu_3, \dots , \mu_M) \non
&+& 2  v^+(\mu_2) X^+(\mu_1) \sum_{j=3}^{M} \f {(-1)^j}{\mu_1 - \mu_j} 
B_{M-3}^{(j)}(\mu_3, \dots , \mu_M) \non
&+& 4  X^+(\mu_1) X^+(\mu_2) \sum_{j=3}^{M} \f {(-1)^j}{\mu_1 - \mu_j} 
\sum_{k=3}^{M} \f {(-1)^{k+{\Th} (j-k)}}{\mu_2 - \mu_k}
B_{M-4}^{(j , k)}(\mu_3, \dots , \mu_M) \non
\eaa
where $B_{M-4}^{(j , k)}(\mu_3, \dots , \mu_M)$  means that the arguments 
$\mu_j$ and $\mu _k$ are omitted.
Since $v^+(\mu)$ commutes with $X^+(\nu)$, the antisymmetry of 
the right hand side of \Ref{asym12} with respect to $\mu _1$ and 
$\mu _2$ follows. Hence we have demonstrated the lemma.
{\qed}

We proceed now to prove the lemma 3.3.

{\it Proof of lemma 3.3.} In particular, when $M=1$ the expression \Ref{vB} 
is just the anticommutator between $v^+(\l)$ and $v^+(\mu)$. 
Using the recurrence relations \Ref{rr} it is straightforward to 
check that the formula \Ref{vB} holds for $M=2$
\baa \la{vB2} 
v^+(\l) B _2 (\mu_1 , \mu_2) &=& B_2 (\mu_1 , \mu_2) v^+(\l) - 2 
\f {X^+(\l) -X^+(\mu _1)}{\l - \mu _1} \; v^+(\mu _2) \non
&+& 2 \f {X^+(\l) -X^+(\mu _2)}{\l - \mu _2}  \; v^+(\mu _1) \;.
\eaa
Therefore we can proceed to demonstrate the lemma 3.3 by induction.
Assume that the relation \Ref{vB} holds for $B_N$, $M \geq N \geq 2$.
Then we have to show the formula \Ref{vB} is valid for $M+1$. 
We use the recurrence relations \Ref{rr} to write
\baa \la{vBN+1a} 
v^+(\l) B_{M+1} &=& v^+(\l) \left( v^+(\mu_1) B_M + 2 X^+(\mu_1) 
\sum_{j=2}^{M+1} \f {(-1)^j}{\mu_1 - \mu_j} B_{M-1}^{(j)} \right) \non
&=& - v^+(\mu_1) v^+(\l) B_M - 2 \f {X^+(\l)-X^+(\mu_1)}{\l - \mu_1} B_M \non
&+& 2 X^+(\mu_1) \sum_{j=2}^{M+1} \f {(-1)^j}{\mu_1 - \mu_j}  
v^+(\l) B_{M-1}^{(j)} \;.
\eaa
Now we can substitute the expressions for $v^+(\l) B_M$ and 
$v^+(\l) B_{M-1}^{(j)}$. After rearranging the terms in an
appropriate way we have
\baa \la{vBN+1b} 
v^+(\l) B_{M+1} &=& (-1)^{M+1} B_{M+1} v^+(\l) + 2 \sum_{j=1}^{M+1} (-1)^j
\f {X^+(\l) -X^+(\mu _j)}{\l - \mu _j} \; B_{M}^{(j)} \;. \non
\eaa
This completes the proof of the lemma.
{\qed}

Subsequently we demonstrate the lemma 3.2.

{\it Proof of lemma 3.2.} The first step in this proof is to use the antisymmetry
property lemma 3.1. The second step is to use the defining recurrence relations.
Then we use the lemma 3.3 and finally after some cancellations we get
the right hand side of the equation \Ref{rrop}.  
\baan
\lefteqn{B_{M}(\mu_1, \dots , \mu_M) \overset{(3.22)}{=}
(-1)^{M-1} B_{M}(\mu_M, \mu_1, \dots , \mu_{M-1}) \overset{(3.20)}{=}
(-1)^{M-1}  v^+(\mu_M) \times}
\\
&&\times B_{M-1}(\mu_1, \dots , \mu_{M-1})
+ (-1)^{M-1} 2 X^+(\mu_M) \sum_{j=1}^{M-1} 
\f {(-1)^{j-1}}{\mu_M - \mu_j} B_{M-2}^{(j)}(\mu_1, \dots , \mu_{M-1})
\\
&&\overset{(3.24)}{=} B_{M-1}(\mu_1, \dots , \mu_{M-1}) v^+(\mu_M) 
+ (-1)^{M-1} 2 \sum_{j=1}^{M-1} (-1)^j \f {X^+(\mu_M) -X^+(\mu _j)}
{\mu_M - \mu _j}  \times 
\\
&&\times B_{M-2}^{(j)} (\mu_1, \dots , \mu_{M-1})
+ (-1)^{M-1} 2 X^+(\mu_M) \sum_{j=1}^{M-1} \f {(-1)^{j-1}}{\mu_M - \mu_j} 
B_{M-2}^{(j)}(\mu_1, \dots , \mu_{M-1}) \non
&&= B_{M-1}(\mu_1, \dots , \mu_{M-1}) v^+(\mu_M) + 2 \sum_{j=1}^{M-1}
(-1)^{M-j-1} \f {X^+(\mu _j)} {\mu_j - \mu _M}  B_{M-2}^{(j)}(\mu_1, \dots , \mu_{M-1})
\eaan
{\qed}

The proofs of the other lemmas are analogous to the proofs of the lemmas 3.1
and 3.3. Since these proofs do not contain illuminating insights and are 
considerably longer than the two we have seen we will omit them.

The recurrence relation \Ref{rr} can be solved explicitly. To be able 
to express the solution of the recurrence relation in a compact form 
it is useful to introduce  a contraction operator $d$.
\begin{definition} Let $d$ be a contraction operator whose action 
on an ordered product $\overset{M}{\underset{\longrightarrow}
{\underset{j = 1}{\prod}}} v^+(\mu_j)$, $M \geq 2$, is given by 
\be \la{d}  
d \left(v^+(\mu_1) v^+(\mu_2) \dots  v^+(\mu_M) \right) = 
2 \sum_{j = 1}^{M-1} X^+(\mu_j) {\sum_{k=j+1}^{M}} 
\f{(-1)^{\sigma(jk)}}{\mu_j - \mu_k} 
\underset{\longrightarrow}{\prod_{m \neq j, k}^{M}} v^+(\mu_m) , 
\ee 
where $\s(jk) $ is the parity of the permutation 
$$ 
\s : (1, 2, \dots , j, j+1,  \dots , k, \dots , M) \to 
(1, 2, \dots , j, k, j+1, \dots , M) \;.
$$ 
\end{definition}
The $d$ operator can be applied on an ordered product
$\overset{M}{\underset{\longrightarrow}{\underset{j = 1}{\prod}}}
v^+(\mu_j)$ consecutively several times. As an illustration, we explicitly apply 
the contraction operator $d$ two times in the case when $M=4$
\baan 
&&d^2 \left(v^+(\mu_1) v^+(\mu_2) v^+(\mu_3) v^+(\mu_4) \right) =
d \left( d \left(v^+(\mu_1) v^+(\mu_2) v^+(\mu_3) v^+(\mu_4) \right) \right) 
\non
&=& 2 X ^+{(\mu _1)} \left( \f{d \left( v^+(\mu_3) v^+(\mu_4) \right)}{\mu _1-\mu _2}
- \f{d \left( v^+(\mu_2) v^+(\mu_4) \right)}{\mu _1-\mu _3} 
+ \f{d \left( v^+(\mu_2) v^+(\mu_3) \right)}{\mu _1-\mu _4} \right) 
\non
&+& 2 X ^+{(\mu _2)} \left( \f{d \left( v^+(\mu_1) v^+(\mu_4) \right)}{\mu _2-\mu _3}
- \f{d \left( v^+(\mu_1) v^+(\mu_2) \right)}{\mu _2-\mu _4} \right) 
+ 2 X ^+{(\mu _3)} \f{d \left( v^+(\mu_1) v^+(\mu_2) \right)}{\mu _3-\mu _4}
\non
&=& 8 X^+(\mu_1) \left( \f{X^+(\mu_3)}{(\mu _1-\mu _2)(\mu _3-\mu _4)}
- \f{X^+(\mu_2)}{(\mu _1-\mu _3)(\mu _2-\mu _4)}
+ \f{X^+(\mu_2)}{(\mu _1-\mu _3)(\mu _2-\mu _4)} \right) 
\eaan
It follows that the contraction operator $d$ can be applied on an ordered 
product\break 
$\overset{M}{\underset{\longrightarrow}{\underset{j = 1}{\prod}}}
v^+(\mu_j)$ up to $[M/2]$ times consecutively. The symbol $[M/2]$ denotes 
the integer part of $M/2$.
\begin{theorem}
Explicit solution to the recurrence relation \Ref{rr} is given by
\be \la{srr}  
B_{M}(\mu_1, \dots , \mu_M) = \underset{\longrightarrow}{\prod_{j=1}^{M}}
v^+(\mu_j) + \sum_{m = 1}^{[M/2]} \f{1}{m!} \; d^m
\underset{\longrightarrow}{\prod_{j = 1}^{M}} v^+(\mu_j) \;,  
\ee
here $d$ is the contraction operator defined above \Ref{d}.
\end{theorem}

The properties of the creation operators $B_M$ studied in this Section
will be\break\hfil 
fundamental tools in determining some of the most important 
characteristics of the $osp(1|2)$ Gaudin model. Our primary interest
is to obtain the spectrum and the eigenvectors of the generating function 
of integrals of motion $t(\l)$ \Ref{tG}.


\section{Spectrum of the $OSp(1|2)$ Gaudin model and its modifications}

With the help of the creation operators $B_M$ it is possible to obtain
the eigenvectors as well as the corresponding eigenvalues of the Gaudin
hamiltonians. This result is a direct consequence of the following theorem.

\begin{theorem}
The lowest spin vector $\Om _-$ \Ref{vac} is an eigenvector
of the generating function of integrals of motion $t(\l)$ \Ref{tG}
with the corresponding eigenvalue $\L _0 (\l)$
\be
\la{l0-t} 
t(\l) \, \Om _- = \L _0 (\l) \, \Om _- \;, \quad \L _0 (\l)= 
\rho^2(\l) + \rho'(\l) \;.   
\ee
Furthermore, the action of the $B$-operators on the lowest spin 
vector $\Om _-$ yields the eigenvectors
\be \la{eigv}  
\Psi (\mu_1, \dots , \mu_M) = B_{M}(\mu_1, \dots , \mu_M) \; \Omega _- \;, 
\ee
of the $t(\l)$ operator 
\be \la{eigeq}  
t(\l) \Psi (\mu_1, \dots , \mu_M) = 
\L (\l ; \, \{\mu_j\}_{j=1}^M) \, \Psi (\mu_1, \dots , \mu_M) \;, 
\ee  
with the eigenvalues
\be \la{l-t} 
\L (\l ; \, \{\mu_j\}_{j=1}^M) = \L _0 (\l) + 
2 \rho (\l) \sum _{k=1}^M \f 1{\l - \mu_k} + 
2 \sum_{k < l} \f{1}{(\l - \mu_k)(\l - \mu_l)} \;, \\
\ee
provided that the Bethe equations are imposed on the parameters
$\{\mu_j\}_{j=1}^M$
\be \la{Beq} 
\b _M (\mu_j) = \rho (\mu_j) + \sum_{k \neq j}^M \f {1}{\mu_j - \mu_k}
= 0 \;.   
\ee 
\end{theorem}
{\it Proof.} The equation \Ref{l0-t} can be checked by a direct substitution 
of the definitions of the operator $t(\l)$ and the lowest spin vector $\Om _-$, 
the equations \Ref{tG} and \Ref{vac}, respectively. 

To show the second part of the theorem, we use the equation \Ref{eigv}
to express the Bethe vectors $\Psi (\mu_1, \dots , \mu_M)$
\be \la{peeq}  
t(\l) \Psi (\mu_1, \dots , \mu_M) = t(\l) \; B_{M}(\mu_1, \dots , \mu_M) \; \Omega _- 
\;. 
\ee
Our next step is to use the lemma 3.6. and the definition of
the lowest spin vector $\Om _-$ the equation \Ref{vac} in order
to calculate the action of the operator $t(\l)$ on the Bethe vectors
when the Bethe equations \Ref{Beq} are imposed 
\be \la{preeq} 
t(\l)B_{M} \Om _- = B_{M} t(\l) \Om _- + 2 \left( \rho (\l) \sum _{i=1}^M 
\f 1{\l -\mu _i} + \sum_{i < j}^M 
\f 1{(\l -\mu _i)(\l -\mu _j)} \right)  B_{M} \Om _- \;.
\ee
We can express the first term on the right hand side since we know how
the operator $t(\l)$ acts on the vector $\Om _-$, the equation \Ref{l0-t},
\be \la{proeeq} 
t(\l) B_{M} \Om _- = \left(\L _0 (\l) + 2 \left( \rho (\l) \sum _{i=1}^M 
\f 1{\l -\mu _i} + \sum_{i < j}^M 
\f 1{(\l -\mu _i)(\l -\mu _j)} \right) \right) B_{M} \Om _- \;.
\ee
The eigenvalue equation \Ref{eigeq} as well as the expression for the eigenvalues
\Ref{l-t} follow from the equation \Ref{proeeq}. 
{\qed}

\begin{corollary} \
In the Gaudin realization of the loop superalgebra given by the \break\hfil
equations \Ref{Gr} and \Ref{Gre} the Bethe vectors $\Psi (\mu_1, \dots , \mu_M)$ 
\Ref{eigv} are the eigenvectors of the Gaudin hamiltonians \Ref{sGh}
\be \la{Geeq}  
H^{(a)} \Psi (\mu_1, \dots , \mu_M) = E^{(a)}_M \Psi (\mu_1, \dots , \mu_M) \;, 
\ee
with the eigenvalues
\be \la{GEn}
E ^{(a)}_M = \underset{b\neq a}{\sum_{b=1}^{N}} \f{l_a \, l_b}{z _a-z _b} 
+ \sum _{j=1}^M  \f{l_a}{\mu _j - z _a} \;,
\ee 
when the Bethe equations are imposed
\be \la{BeqG} 
\b _M (\mu_j) = \rho (\mu_j) + \sum_{k \neq j}^M \frac {1}{\mu_j - \mu_k} = \sum_{a=1}^N 
\frac {-l_a}{\mu_j - z_a} + \sum_{k \neq j}^M \frac {1}{\mu_j - \mu_k} = 0 \,.   
\ee 
\end{corollary}
{\it Proof.} The statement of the corollary follows from residue of the
equation \Ref{eigeq} at the point $\l = z _a$. The residue can be
determined using \Ref{tpole}, \Ref{l-t} and \Ref{l0-t}.
{\qed}

The eigenvalue \Ref{l-t} of the operator $t(\l)$ and the Bethe equations
\Ref{Beq} can be obtained also as the appropriate terms in the 
quasi-classical limit $\eta \to 0$ of the expressions \Ref{lq} and
\Ref{BEq}.

Comparing the eigenvalues $E^{(a)}_M$ \Ref{GEn} of the Gaudin hamiltonians
and the Bethe equations \Ref{BeqG} with the corresponding quantities of
the $sl(2)$ Gaudin model \c{GB,S87} we arrive to an interesting observation.

\begin{remark}
The spectrum of the $osp(1|2)$ Gaudin model with the spins $l_a$ coincides 
with the spectrum of the $sl(2)$ Gaudin system for the integer spins 
(cf. an analogous observation for partition functions of corresponding 
anisotropic vertex models in \c{HS}). 
\end{remark}

\begin{remark}
The Bethe vectors are eigenstates of the global generator  $h _{gl}$
\be \la{hglee}
h _{gl} \Psi (\mu_1, \dots , \mu_M) = 
\left( - \sum _{a=1}^N l_a  + M \right) \Psi (\mu_1, \dots , \mu_M)
\;.
\ee
Moreover, these Bethe vectors are the lowest spin vectors
of the global $osp(1|2)$ since they are annihilated by the 
generator $v^-_{gl}$
\be \la{vglBv} 
v^-_{gl} \Psi (\mu_1, \dots , \mu_M) = 0 \;, 
\ee
once the Bethe equations are imposed \Ref{BeqG}. These conclusions follow
from the remark 3.2 the equations \Ref{hglB}, \Ref{v-glB} and the definition
of the Bethe vectors \Ref{eigv}.  
\end{remark}

Hence, action of the global generator $v^+_{gl}$ on the lowest spin
vectors $\Psi (\mu_1, \dots , \mu_M)$ generates a multiplet of
eigenvectors of the operator $t(\l)$
\be \la{mult}
\left(v^+_{gl}\right) ^m \Psi (\mu_1, \dots , \mu_M) \;, 
\quad m=1, 2, \ldots , 2 \left(\sum _{a=1}^N l_a  - M\right) \;. 
\ee
One can repeat the arguments of \c{TF,F95} to demonstrate combinatorially
completeness of the constructed states.

As was pointed out already in \c{GB} for the $sl(2)$ case,
there are several modifications of the hamiltonians (\ref{sGh}). One of them
is the Richardson's pairing-force hamiltonian \c{Rich}. 
These modifications can be formulated in the framework of universal
L-operator and $r$-matrix formalism (\ref{rL}) \c{S87}.

Due to invariance of the $r$-matrix \Ref{cra}
\baa \la{r-inv}
\left[ r (\l) , Y \otimes I + I \otimes Y \right] = 0 \, , 
\quad Y \in osp (1|2)
\eaa
one can add to the L-operator any element of $osp(1|2)$
\baa \la{L-mod}
L (\l) \to \tilde{L} (\l) = g \, Y +  L (\l) \, ,
\eaa
preserving commutation relations \Ref{rL}. If we choose $Y = h$, then 
\baa \la{t-mod}
\wtt (\l) = \half \, {\str} \, \widetilde{L} ^2(\l) = t (\l) 
+ 2 g \, h (\l) + g ^2\, ,
\eaa
will have the commutativity property, {\ie} $\wtt (\l) \wtt (\mu)
= \wtt (\mu) \wtt (\l)$. Hence we can take $\widetilde{t} (\l)$
to be the generating function of the (modified) integrals of motion
\baa \la{tpole-mod}
\wtt (\l) &=& \sum_{a =1}^N \left( \f {c_2(a)}{(\l - z_a)^2} 
+ \, 2 \, \f {\widetilde{H}^{(a)}}{\l - z_a} \right) + g ^2\, , \\ 
\la{h-mod}
\widetilde{H}^{(a)} &=& g \, h _a + \sum_{b \neq a} 
\frac {c_2^{\otimes}(a,b)}{z_a - z_b} \, .  
\eaa
Notice that the global $osp(1|2)$ symmetry is now broken down to global
$u(1)$ 
\baa \la{tmod-sym}
\left[ \wtt (\l), h _{gl} \right] &=& \left[ \wtt (\l) \; , \; 
\sum _{a=1}^N \; h _a \right] = 0 \, .
\eaa
In this case the eigenstates $\Psi _M$ are generated by the same
B-operators. However, corresponding eigenvalues and Bethe equations
are now given by
\baa \la{l-t-mod}
\widetilde{\L} (\l ; \, \{\mu_j\}_{j=1}^M) &=& \L (\l ; \, \{\mu_j\}_{j=1}^M)
+ 2 g \, \rho (\l) + 2 g \, \sum_{j=1}^M  \frac 1{\l - \mu_j} + g ^2 \, , \\
\la{evGh-mod}
\widetilde{E} ^{(a)}_M &=& E ^{(a)}_M + g \, ( -l _a) \, , \\
\la{Beq-mod}
- g + \sum_{a=1}^N \f {l_a}{\mu_j - z_a} &=& \sum_{k \neq j}^M 
\f 1{\mu_j - \mu_k} \,.
\eaa 
The crucial step in the proof of these equations is the observation
that the commutation relations between the operator $\wtt (\l)$ \Ref{t-mod}
and the creation operators $B_M$ are equal to the commutation relations 
\Ref{tB} but with modified operator $\hb _M(\mu_j) \to \hb _M(\mu_j) + g$.
To see this notice the similarity between the terms with $v ^+(\l) B ^{(i)}_{M-1}$ 
operators and with $X ^+(\l) B ^{(i,j)}_{M-2}$ operators 
in the lemma 3.4 the equation \Ref{hB} and in the lemma 3.6 the equation \Ref{tB}.

Richardson like hamiltonian \c{Rich} can be obtained as a coefficient
in the $\l \to \infty$ expansion
\baa \la{t-mod-inf}
\wtt (\l) &=& g ^2 + \f {2g}{\l} \, \sum _{a=1}^N h _a +  
\f 1{\l ^2} \, \left( 2g \, \sum _{a=1}^N z_a h _a + c _2 (gl) \right) 
+ O ( \f 1{\l ^3}) \, .
\eaa
The first coefficient in this expansion is $h _{gl}$, global $u(1)$ symmetry
generator (\ref{tmod-sym}). Let us denote the second coefficient by $2H _g$.
Using the global $osp (1|2)$ algebra generators one can write 
this hamiltonian in the form
\baa \la{hRich}
H _g &=& g \, \sum _{a=1}^N z_a h _a + 2 X ^+_{gl} X ^-_{gl} 
+ v ^+_{gl} v ^-_{gl} \, ,
\eaa
where we have omitted the term $h _{gl} (h _{gl} -1)$. 
The eigenvalues of the hamiltonian $H _g$ are given by 
\be \la{HR-ev}
H _g \Psi_M(\mu_1, \ldots, \mu_M) = E_g \Psi_M(\mu_1, \ldots, \mu_M) \;,
\quad  E_g = \sum _{j=1}^M \mu_j - \sum_{a=1}^N l_a z_a \;.
\ee

A realization of the generators in terms of fermionic oscillators 
in the $sl(2)$ case yields Richardson hamiltonian \c{Rich}. In the 
$osp (1|2)$ case, which we consider here, there are extra fermionic degrees of 
freedom due to the term $v ^+_{gl} v ^-_{gl}$ and constraints
$(v ^{\pm}_{gl}) ^2 = X ^{\pm}$.

One can realize Sklyanin bracket \Ref{rL} using an L-operator
with bosonic and fermionic oscillator entries
\be \la{L-osc}  
L _{osc}(\l) = \left( \begin{array}{ccc}
\l & - \g & 2 b \\
\g ^+ & 0 & \g  \\
2b ^+ & \g ^+ & - \l 
\end{array} \right) \; , 
\ee 
where
\baa \la{osc}
\left[ b \, , \, b ^+ \right] = 1 \;, \quad
\left[ \g \, , \, \g ^+ \right] _+ = 1 \;, \quad 
\g ^2 = (\g^+ ) ^2 = 0 \;.
\eaa
It is straightforward to see that the corresponding realization of the 
loop superalgebra will have only two nonzero commutators. 
Hence, one can consider a combination of Gaudin and oscillator 
realizations
\baa \la{L-comb}
\widetilde{L} (\l) = L _{osc}(\l) + L (\l) \;.
\eaa
Intergals of motion can be obtained using
\baa \la{t-comb}
\widetilde{t} (\l) &=& \half \, {\str} \left(  L ^2_{osc}(\l) + 
2 L _{osc}(\l) L (\l) +  L ^2 (\l) \right) 
\nonumber\\
&=& t (\l) + \l ^2 + 2 (b \, b^+ + b^+b) + (\g^+ \g - \g \, \g ^+)
\nonumber\\
&+& 2 \left( \l h (\l) + 2 \left( b X ^+ (\l) +  b ^+ X ^- (\l) \right) +
\left( \g^+ v ^- (\l) -  \g v ^+ (\l) \right) \right) \;.
\eaa
Correspoding B-operators can be constructed using 
\baa \la{vX-comb}
\widetilde{v} ^+ (\l) =  \g^+ + v ^+ (\l) \; , && 
\widetilde{X} ^+ (\l) =   b^+ + X ^+ (\l) \; .
\eaa
Finaly, the eigenvalues $\widetilde{\L}$ and the Bethe equations
are given by
\baa \la{l-t-comb}
\widetilde{\L} (\l ; \, \{\mu_j\}_{j=1}^M) &=& \L (\l ; \, \{\mu_j\}_{j=1}^M)
+ 2 (\l \, \rho (\l) + 1 ) + \sum_{j=1}^M  \frac  {2 \l}{\l - \mu_j} + \l ^2 \\
\la{Beq-comb}
- \mu _j + \sum_{a=1}^N \f {l_a}{\mu_j - z_a} &=& \sum_{k \neq j}^M 
\f 1{\mu_j - \mu_k} \,.
\eaa

Further modifications can be obtained considering quasi-classical 
limit of the quantum spin system with non-periodic boundary conditions 
and corresponding reflection equation.

The expression of the eigenvector of a solvable model in terms of 
local variables parameterized by sites of the chain or by space 
coordinates is known as coordinate Bethe Ansatz \c{GB}. The coordinate 
representation of the Bethe vectors gives explicitly analytical 
dependece on the parameters $\{ \mu _i\}_1^M$ and $\{ z_a\}_1^N$
useful in a relation to the  Knizhnik-Zamolodchikov equation (Section 5).
Using the Gaudin realization \Ref{Gr} of the generators 
\ben
v^+(\mu) =  \sum_{a=1}^N \f {v^+_a}{\mu - z_a} \;, \quad 
X^+(\mu) =  \sum_{a=1}^N \f {X^+_a}{\mu - z_a} \;,  
\een
and the definition of the creation operators \Ref{srr},
one can get the coordinate representation of the $B$-operators: 
\be \la{coorB}  
B_{M}(\mu_1, \mu_2, ..., \mu_M) = \sum_{\pi} \left(v^+_{a_1}
\cdots v^+_{a_M} \right) _{\pi} \prod_{a=1}^N 
\vph (\{ \mu_m^{(a)}\}^{\mid \Ka _a\mid}_1 ; z_a ) \;,  
\ee
where the first sum is taken over ordered partitions $\pi$ 
of the set $(1, 2, \ldots , M)$ into subsets $\Ka _a$, 
$a= 1, 2, \ldots , N$, including empty subsets with the constraints
\ben
\bigcup_a \Ka _a = (1, 2, \ldots , M) \;, \quad 
\Ka _a \bigcap \Ka _b = \emptyset \quad \hbox{for} \; a \neq b \;.
\een
The corresponding subset of quasimomenta
\ben
\left( \mu _1^{(a)} = \mu _{j_1} , \mu _2^{(a)} = \mu _{j_2} ,
\ldots \mu _{\mid \Ka _a \mid}^{(a)} = \mu _{j_{\mid \Ka _a\mid }} 
; j _m \in \Ka _a \right) \;,
\een
where $\mid \Ka _a\mid$ is the cardinality of the subset $\Ka _a$,
and $j_k<j_{k+1}$, entering into the coordinate wave function
\ben
\vph (\{ \nu _m \}_1^{\mid \Ka \mid} ; z) = 
\sum _{\s \in {\cal S}_{\mid \Ka \mid}} (-1)^{p(\s)}
\left( (\nu_{\s(1)} - \nu_{\s(2)})  (\nu_{\s(2)} - \nu_{\s(3)}) \cdots
(\nu_{\mid \Ka \mid} - z) \right)^{-1} \;.  
\een
Due to the alternative sum over permutations 
$\s \in {\cal S}_{\mid \Ka \mid}$ this functions is
antisymmetric with respect to the quasi-momenta. Finally the first 
factor in \Ref{coorB}
\ben
\left(v^+_{a_1} \cdots v^+_{a_M} \right) _{\pi}
\een
means that for $j_m \in \Ka _a$ corresponding indices of $v^+_{a_{j_m}}$
are equal to $a$ so that  $v^+_{a_{j_m}} = v^+_a$. One can collect these
operators into product $\prod _{a=1}^N \left(  v^+_a \right) ^{\mid \Ka _a \mid}$,
consequently we have an extra sign factor $(-1) ^{p(\pi)}$.

This coordinate representation is similar to the representations 
obtained in \c{BF,FFR,ReshVar} for the Gaudin models related to 
the simple Lie algebras (see also \c{M97}). The $Z_2$-grading results 
in extra signs, while the complicated structure 
of the $B_M$-operators (for the $sl(2)$-GM they are just products 
of $B_1$-operators $B_1(\mu_j) = X ^+(\mu_j)$ ) is connected with 
the fact that $(v^+_j)^2 = X_j^+ \neq 0$ while for $j \neq k$ $v^+_j$ 
and $v^+_k$ anticommute. 


\section{Solutions to the Knizhnik-Zamolodchikov equation}

Correlation functions $\psi (z _1 , \dots , z_n)$ of a two dimensional
conformal field theory satisfy the Knizhnik-Zamolodchikov equation 
\c{KZ}
\baa \la{KZ}
\ka \f {\p}{\p z_a} \psi (z _1 , \dots , z_n) &=&
\left(\sum _{b\neq a} \f {Y^{\a}_a \otimes  Y^{\a}_b}{z_a - z_b} \right)
\psi (z _1 , \dots , z_n) \;,
\eaa
where $Y^{\a}_a$ are generators of an orthonormal basis of a simple
Lie algebra in a finite dimensional irreducible representation $V_a$ and 
$\psi (z _1 , \dots , z_n)$
is a function of $N$ complex variables taking values in a tensor product
$\underset {a=1}{\overset {N}{\otimes}} V_a$. The operator on the right hand 
side of \Ref{KZ} is a Gaudin hamiltonian \Ref{Gh}.

A relation between the Bethe vectors of the Gaudin model related to 
simple Lie algebras and the solutions to the
Knizhnik-Zamolodchikov equation is well known for
sometime \c{BF, FFR}. Approach used here to obtain solutions to the
Knizhnik-Zamolodchikov equation corresponding to super-conformal field 
theory and  Lie superalgebra $osp(1|2)$ starting from B-vectors \Ref{eigv} 
is based on \c{BF}.

A solution in question is represented as a contour integral over the variables
$\mu_1 \dots \mu_M$
\baa \la{KZ-sol}
\psi ( z_1, \dots z_N ) &=& \oint \dots \oint \; \ph ( \vm | \vz ) \Psi ( \vm | \vz )
\; d\mu _1 \dots d\mu _M \; , 
\eaa
where an integrating factor $\ph ( \vm | \vz )$ is a scalar function
\baa \la{factor}
\ph ( \vm | \vz ) &=& \prod _{i< j}^M ( \mu _i - \mu _j ) ^{\f 1{\ka}}
\prod _{a< b}^N ( z _a - z _b ) ^{\f {l_al_b}{\ka}}
\left( \prod _{k=1}^M \prod _{c=1}^N ( \mu _k - z _c ) ^{\f {-l_c}{\ka}}
\right) \;, 
\eaa
and $\Psi ( \vm | \vz )$ is a Bethe vector \Ref{eigv} where the corresponding
Bethe equations are not imposed. 
 
As a first step in the proof that $\psi ( z_1, \dots z_N )$ given by \Ref{KZ-sol}
is a solution of \Ref{KZ} we differentiate the product $\ph \Psi$ with 
respect to $z_a$ and obtain
\baa \la{calc}
\p _{z_a } \left( \ph  \Psi \right) = \p _{z_a } \left( \ph  \right ) \Psi +
\ph  \p _{z_a } \left( \Psi \right ) \;.
\eaa
Using \Ref{factor} the first term on the right hand side can be calculated explicitly 
\baa
\ka \p _{z_a } \ph = \left(\underset{b\neq a}{\sum _{b=1}^N}
\f {l_a \, l_b}{z _a-z _b} - \sum _{j=1}^M  \f {l_a}{z _a - \mu _j} \right) \ph 
= E ^{(a)}_M \ph \;.
\eaa
Furthermore, taking a residue of \Ref{tB} at $\l = z_a$ we have
\be
H ^{(a)} \Psi =  E ^{(a)}_M \Psi + \sum _{j=1}^M \f {(-1) ^j}{z_a-\mu _j}
\b _M ( \mu _j ) {\tPsi} ^{(j,a)} \;,
\ee
where
\be
{\tPsi} ^{(j,a)} = \left( v ^+_a B ^{(j)}_{M-1} + 2 X ^+_a \sum_{k\neq j}^M 
\f{(-1)^{k + \Th(j-k)}}{\mu _j - \mu _k} B ^{(j,k)}_{M-2} \right) \; \Om _- \;. 
\ee
Hence \Ref{calc} can be written as
\be
\ka \p _{z_a } \left( \ph  \Psi \right) = H ^{(a)} \left( \ph  \Psi \right) +
\ph \sum _{j=1}^M \f {(-1) ^{j}}{\mu _j - z_a} \b _M ( \mu _j ) 
{\tPsi} ^{(j,a)} + \ka \ph  \p _{z_a } \left( \Psi \right )\;.
\ee
Moreover, from \Ref{factor} we also have
\be \la{der1}
\ka \p _{\mu_j } \ph = \left(\sum _{a=1} ^N \f {-l_a}{\mu _j - z _a} 
+  \mathop{\sum _{j=1}} \limits_{j\neq k}\limits^M  
\f 1{\mu_j - \mu _k} \right) \ph = \b  _M (\mu _j) \ph \;,
\ee
and from the lemma 3.7 follows
\be \la{der2}
\p _{z_a } \Psi = \sum _{j=1}^M (-1) ^{j} \p _{\mu_j } 
\left ( \f {{\tPsi} ^{(j,a)}}{\mu _j - z_a} \right )
\ee
Thus, using \Ref{der1} and \Ref{der2},  
\be \la{final}
\ka \p _{z_a } \left( \ph  \Psi \right) = H ^{(a)} \left( \ph  \Psi \right) +
\ka \sum _{j=1}^M \p _{\mu_j } \left( \f {(-1) ^{j}}{\mu _j - z_a} \;
\ph \, {\tPsi} ^{(j,a)} \right) \;.
\ee
A closed contour integration of $\ph \Psi$ with respect to 
$\mu _1 , \dots ,\mu _M$
will cancel the contribution from the terms under the sum in \Ref{final}
and therefore $\psi ( z_1, \dots z_N )$ given by \Ref{KZ-sol} 
satisfies the Knizhnik-Zamolodchikov equation.

Conjugated Bethe vectors $(B_M \Om_-) ^{\ast}$ are entering into the
solution $\tpsi( z_1, \dots z_N )$ of the dual Knizhnik-Zamolodchikov equation
\baa \la{dKZ}
- \ka \f {\p}{\p z_a} \tpsi (z _1 , \dots , z_N) &=& 
\tpsi (z _1 , \dots , z_N) \; H ^{(a)} \;.
\eaa
The scalar product $\left( \tpsi (z _1 , \dots , z_N) \; , 
\; \psi (z _1 , \dots , z_N) \right)$ does not depend on $\{z_j\} _1^N$
and its quasi-classical limit $\ka \to 0$ gives the norm of the Bethe
vectors due to the fact that the stationary points of the contour integrals 
for $\ka \to 0$ are solutions to the Bethe equations \c{ReshVar}
\baa \la{KZBeq}
\f{\p S}{\p \mu_j} &=& \sum _{a=1} ^N \f {-l_a}{\mu _j - z _a} 
+  \mathop{\sum _{j=1}} \limits_{j\neq k}\limits^M  
\f {1}{\mu_j - \mu _k} = 0 \;, \\
\la{S}
S ( \vm | \vz ) &=& \ka \ln \ph = \sum_{a < b}^N l_a l_b \ln (z_a - z_b)
+ \sum _{i < j}^M \ln ( \mu _i - \mu _j) - \sum _{a=1}^N \sum _{j=1}^M
l _a \ln (z_a - \mu _j) \;. \non
\eaa
According to the remark in the end of Section 4 analytical properties of 
the Bethe vectors of the $osp(1|2)$ Gaudin model coincide with the analytical 
properties of the $sl(2)$ Gaudin model. Thus, the expression for the norm of the
Bethe vectors $\Psi$ \Ref{eigv} obtained as the first term in the asymptotic
expansion $\ka \to 0$ coincides also
\baa \la{norm}
&&\left( \Psi \; , \; \Psi \right) =  \det \left(
\f {\p ^2 S}{\p \mu_j \; \p \mu_k} \right) \;, \\
\f {\p ^2 S}{\p \mu_j^2} &=& \sum _{a=1}^N \f{l_a}{(\mu _j - z_a)^2} - 
\sum _{k \neq j}^M \f{1}{(\mu _j - \mu _k)^2} \;, 
\quad \f{\p ^2 S}{\p \mu_j \; \p \mu_k} = \f{1}{(\mu _j - \mu _k)^2} \;, 
\quad \hbox{for $j\neq k$} \;. \non
\eaa
Finally we notice that the modification of the Gaudin hamiltonians we discussed
at the end of the previous Section can be easily transfered to the corresponding
modification of the Knizhnik-Zamolodchikov equations. The modification \Ref{L-mod} 
for the $sl(2)$ 
Gaudin model was studied in \c{BK} as a quantization of the Schlesinger system 
(see also \c{R92}). Both modifications are related with extra factors in 
the integrating scalar function \Ref{factor}
\be \la{f-i}
\ph _j = \exp \left(\f{S_j}{\ka}\right) \;, \quad j=0 , 1, 2\;,
\ee
where $S_0 = S$ \Ref{S} and
\baa 
\la{S-1}
S _1 &=& S _0 + g \sum _{j=1}^M \mu _j - g \sum _{a=1}^N l_a z_a \;, \\
\la{S-2}
S _2 &=& S _0 + \f{1}{2} \sum _{j=1}^M \mu _j^2 - \f{1}{2} \sum _{a=1}^N l_a z_a^2 \;,
\eaa
correspond to the first \Ref{L-mod} and second \Ref{L-comb} modification
respectively.


\section{Conclusion} 

The Gaudin model corresponding to the simplest non-trivial Lie superalgebra
$osp(1|2)$ was studied. A striking similarity between some of the most 
fundamental characteristics of this system and the $sl(2)$ Gaudin model 
was found. Although explicitly constructed creation operators $B_M$
\Ref{srr} of the Bethe vectors are complicated polynomials of the
$L$-operator entries $v^+(\l)$ and $X^+(\l)$, the coordinate form of 
the eigenfunctions differs only in signs from the corresponding states in 
the case of $sl(2)$ model. Moreover, the eigenvalues and the Bethe equations 
coincide, provided that the $sl(2)$ Gaudin model with integer spins is 
considered. 

Let us point out that by the method proposed in this paper one can 
construct explicitly creation operators of the Gaudin models 
related to trigonometric Izergin-Korepin $r$-matrix \c{T} 
and trigonometric $osp(1|2)$ $r$-matrix \c{LS} which have the
same matrix structure as \Ref{r0}. 
Similarly to the simple Lie algebra case solutions to the 
Knizhnik-Zamolodchikov equation were constructed from the Bethe vectors 
using algebraic properties of the creation operators $B_M$ and the 
Gaudin realization of the loop superalgebra ${\cal L}_+(osp(1|2))$. 
This interplay between the Gaudin model and the Knizhnik-Zamolodchikov
equation enabled us to determine the norm of eigenfunctions of the 
Gaudin hamiltonians
\ben 
\parallel \Psi (\mu _1 , \dots \mu _M ; \{ z_a\}_1^N ) \parallel ^2 =
\det \left( \f{\p ^2 S}{\p \mu_j \; \p \mu_k} \right) \;.
\een
The difficult problem of correlation function calculation
for general Bethe vectors
\ben
{\cal C} \left( \{ \nu_j \}_1^M ; \{ \mu _i\}_1^M ; \{ \l_k\}_1^K \right) =
\left( \Om _- \; , \; B_M ^{\ast} (\nu _1 , \dots \nu _M ) \prod_{k=1}^K
h(\l_k) \, B_M (\mu _1 , \dots \mu _M ) \Om _- \right)
\een
was solved nicely for the $sl(2)$ Gaudin model in \c{S99} using the
Gauss factorization of the loop algebra group element and the
appropriate Riemann-Hilbert problem. Although the corresponding 
factorization is known even for the quantum superalgebra $osp_q(1|2)$ 
\c{DKS} the final expression of the correlation functions is
difficult to obtain due to the complicated structure of the creation
operators $B_M(\mu _1 , \dots \mu _M ) = \hbox{Poly} ( v ^+ , X^+)$ \Ref{srr}.
The study of this problem is in progress and the following expression 
for the scalar product of the Bethe states is conjectured 
(cf. \c{S99})
\ben
\left( \Om _- \; , \; B_M ^{\ast} (\nu _1 , \dots \nu _M ) 
B_M (\mu _1 , \dots \mu _M ) \Om _- \right) = \sum _{\s \in {\cal S}_M} 
(-1) ^{p(\s)} \det {\cal M}^{\s} \;,
\een
where the sum is over symmetric group ${\cal S}_M$ and $M\times M$ matrix
${\cal M}^{\s}$ is given by
\baan
{\cal M}^{\s}_{jj} &=& \f{\r(\mu_j) -\r( \nu _{\s(j)})}{\mu _j - \nu _{\s(j)}}
- \sum_{k\neq j}^M \f{1}{(\mu _j - \mu _k)(\nu _{\s(j)} - \nu _{\s(k)})} \;, \\
{\cal M}^{\s}_{jk} &=& \f{1}{(\mu _j - \mu _k)(\nu _{\s(j)} - \nu _{\s(k)})} \;,
\quad \hbox{for $j,k=1, 2, \ldots M$}.
\eaan

\section{Acknowledgements}

We acknowledge useful discussions and communications with N. ~Yu. ~Reshetikhin,
V. ~O. ~Tarasov and T. ~Takebe. This work was supported by the grant 
PRAXIS XXI/BCC/22204/99, INTAS grant N 99-01459 and FCT project SAPIENS-33858/99.

\end{document}